%% file: kraus.tex
\documentclass[aps,floatfix,prc,twocolumn,showpacs,showkeys,amsmath,amssymb,superscriptaddress]{revtex4}
\usepackage{graphicx}
\usepackage{dcolumn}% Align table columns on decimal point
\usepackage{bm}% bold math
\usepackage{color}
\begin{document}
\title{Particle production in $p-p$ collisions  and prediction for LHC energy}
\author{I.~Kraus}
\affiliation{Institut f\"ur Kernphysik, Darmstadt University of Technology, D-64289 Darmstadt, Germany}
\affiliation{Nikhef, Kruislaan 409, 1098 SJ Amsterdam, The Netherlands}
\author{J.~Cleymans}
\affiliation{Institut f\"ur Kernphysik, Darmstadt University of Technology, D-64289 Darmstadt, Germany}
\affiliation{UCT-CERN Research Centre and Department  of  Physics,\\ University of
 Cape Town, Rondebosch 7701, South Africa}
\author{H.~Oeschler}
\affiliation{Institut f\"ur Kernphysik, Darmstadt University of Technology, D-64289 Darmstadt, Germany}
\author{K.~Redlich}
\affiliation{Institut f\"ur Kernphysik, Darmstadt University of Technology, D-64289 Darmstadt, Germany}
\affiliation{Institute of Theoretical Physics, University of Wroc\l aw, Pl-45204 Wroc\l aw, Poland}
\date{\today}
\begin{abstract}
We analyze  recent data on particle production yields obtained in $p-p$ collisions at SPS and RHIC
energies within the statistical model. We apply the  model formulated in the canonical ensemble and
focus on strange particle production.  We introduce different methods to account for strangeness
suppression effects  and discuss their phenomenological verification. We show that at RHIC  the
midrapidity data on strange and multistrange particle multiplicity can be successfully described by
the canonical statistical model with and without an extra suppression effects. On the other hand,
SPS data integrated over the full phase-space require an additional strangeness suppression factor
that is beyond the conventional canonical model. This factor  is  quantified by the strangeness
saturation parameter  or strangeness  correlation volume. Extrapolating all relevant thermal
parameters from SPS and  RHIC to LHC energy we present predictions of the statistical model for
particle yields in $p-p$ collisions at $\sqrt{s}$ = 14~TeV. We discuss the role and the influence of
a strangeness correlation volume on   particle production in $p-p$ collisions at LHC.
\end{abstract}
\maketitle
\section{\label{secIntroduction}Introduction}
Results from the Large Hadron Collider (LHC) have a potential  to usher in a new era of discoveries
and insights into particle physics. In order to fully appreciate  these it is important to have a
clear understanding  as to what we expect will happen at the LHC without new physics and new phenomena
involved. This requires an extrapolation of trusted models to LHC energies. In this paper we use the
statistical model  for  particle production which has been around since more than half a century~\cite{fermi,heisenberg,hagedorn} and discuss its predictions for LHC energy.

The statistical   model has been very successful in describing hadron yields in central heavy-ion
collisions~\cite{model,kr,anton,biro}. It has provided a useful framework for describing centrality
and system size dependence of particle production in  heavy-ion collisions \cite{nuxu,jpsi}.
It has also led to new insights about its applicability  in small systems like  $p-p$ and even $e^+-e^-$ scattering \cite{kr,becattini,krpbm,hs,small}.

In the following we concentrate on particle production in elementary collisions and discuss the predictions of the statistical model for particle yields in $p-p$ collisions at the LHC at $\sqrt{s}$ = 14~TeV center of mass energy.

Particle production  calculated within the statistical model is quantified by a set of thermal
parameters, these are: the temperature,  the volume  and the set of chemical potentials which are
related to conserved charges. To make any predictions for particle
production at  LHC energies one needs methods to extrapolate the thermal parameters to higher
energies. In this paper we first analyze the recent experimental data
obtained in  $p-p$ collisions at SPS and RHIC  and then we extrapolate these to LHC beam
energies. This allows us to make predictions for different particle yield ratios at the LHC
as well as discuss   their dependence on the values of extrapolated parameters. A very brief account
of some of our results was given in~\cite{lastCall}. In this   paper we present a complete
description of the extrapolation method which is based on our new analysis of the $p-p$  data
at SPS and RHIC energies.

The paper is organized as follows: In Section~\ref{secModel} we summarize the main features of the
canonical statistical  model. In  Section~\ref{secFit} we present the analysis of SPS and
RHIC data obtained in  $p-p$ collisions. In Section~\ref{secPredict} we introduce  the extrapolation
of the model to the LHC energy and discuss its predictions for particle production in $p-p$
collisions. In the final section we present conclusions and summarize our results.
\section{\label{secModel}Statistical Model}
Systematic studies of particle yields extending over more than a decade, using experimental results
at different beam energies, have revealed a clear underlying freeze-out pattern for particle yields
in heavy-ion collisions ~\cite{eovern}. A detailed comparison of different freeze-out criteria was
made in~\cite{wheaton_phd} and we followed the one used previously~\cite{lhcPbPb} to extrapolate the
temperature $T$ and baryon chemical potential $\mu_{B}$  to the LHC energy. The expected hadron
multiplicity ratios  can be calculated directly from this extrapolation if the grand canonical (GC)
description of charge conservation is adequate. This is because, in the GC ensemble, any  particle
multiplicity ratio is uniquely determined by the  values of the  temperature and the chemical
potentials.
\subsection{\label{secCanonical}Canonical suppression}
The usual form of the statistical model in the grand canonical ensemble formalism cannot be used
when  either the temperature or the volume  or both are small, as a rule of thumb one needs $VT^3>1$
for a grand canonical description to hold~\cite{hagedornred,rafeldan}. Furthermore, even if this
condition is met, if the abundance of a subset of particles carrying a conserved charge is small,
the canonical suppression still appears even though  the grand canonical description is valid for
the bulk of the produced hadrons.  There is by now a vast literature on the subject of canonical
suppression and we refer to several  articles (see e.g.~\cite{kr,polishreview,can1}).

\begin{figure}
\includegraphics[width=0.9\linewidth]{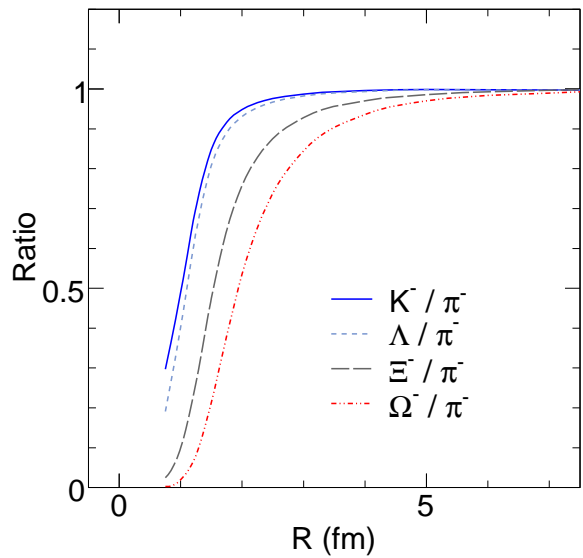}
\caption{\label{fig1} Particle ratios as a function of the volume radius $R$.
The temperature $T$ = 170 MeV and the baryon chemical potential $\mu_B$ = 1 MeV were chosen according to the thermal conditions expected to be valid at LHC energy. All ratios are normalized to their grand-canonical values.}
\end{figure}

The effects of canonical suppression are illustrated in Fig.~\ref{fig1} which shows particle ratios
of strange and multi-strange hadrons to pions normalized to the values in the grand
canonical limit as a function of the radius of the system. The smaller the volume
and the larger strangeness content of the particle the stronger  the suppression of the
yield of strange particles. This has been discussed in great detail
in~\cite{hamieh,small}.

The analysis of the variations of particle ratios with the size of the system, e.g. via the number
of participants, at  SPS revealed~\cite{small} that the experimental data show stronger suppression
of strange-particle yields than that   expected in the   canonical model
~\cite{hotQuarks,small,lisbon}. Consequently, an additional suppression effect had to be
included in order to quantify the observed yields of strange particles.
\subsection{\label{secFitExtra}Strangeness correlation volume}
One possible explanation for the failure of the canonical corrections is that strangeness can be
conserved exactly in a small  subvolume of the fireball, thus leading to a stronger canonical
suppression, as seen in Fig.~\ref{fig1}, even though the strange particles within such subvolume are
taken as being in chemical equilibrium. A modification of the statistical model was formulated
in~\cite{hotQuarks,small} that allows to quantify this extra suppression by the strangeness
correlation volume  (cluster size)  within which the  strangeness is conserved exactly. An
alternative method was proposed in~\cite{becattini} to include in the canonical statistical model an
additional factor $\gamma_S$ that accounts for possible deviations of strange particle abundance
from their chemical equilibrium distribution value.

In the  next section we discuss how to choose  the  size of the cluster or  $\gamma_S$  in the
canonical model to describe    experimental data on strange particle production  in $p-p$
collisions. First, we use data from SPS and  from RHIC energies and   next we discuss the possible
extrapolation of the model parameters to make  predictions for particle production  in $p-p$
collisions at LHC energies.
\begin{figure}
\includegraphics[width=0.99\linewidth]{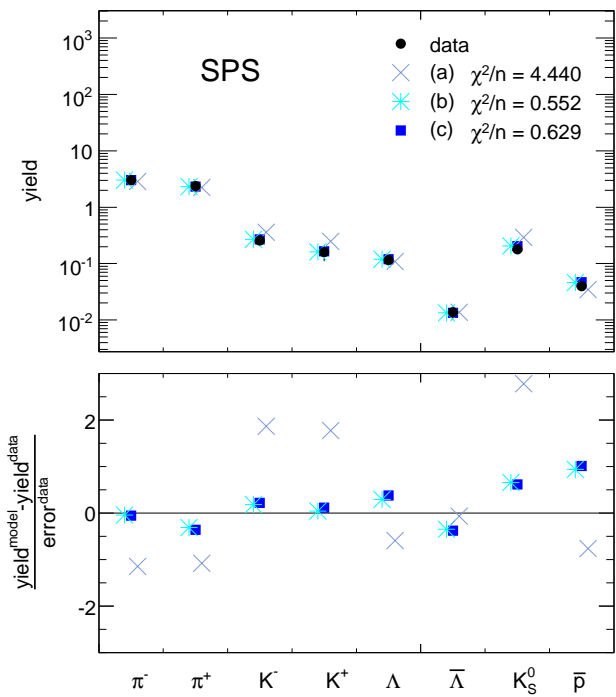}
\caption{\label{fig2} Integrated particle yields from p-p collisions at
$\sqrt{s}$~=~17.3~GeV~\cite{dataSPSpi,dataSPSK,dataSPSlam,dataSPSK0s} together with different model
predictions: (a), (b) and (c) introduced in the text. The lower panel shows the
deviation of the model fits to data.}
\end{figure}
\section{\label{secFit}Data analysis}
At the top SPS energy   data are available for $\pi,~K,~K^0_S,~\bar{p}$ and $\Lambda$ yields
integrated over the full phase-space ~\cite{dataSPSpi,dataSPSK,dataSPSlam,dataSPSK0s} in $p-p$
collisions. In Fig.~\ref{fig2} we show the comparison of the statistical model with these data for
three different implementations of the canonical statistical model labelled by (a), (b) and (c). In
all cases the conservation of the electric charge and baryon number is formulated in the GC ensemble
and is thus controlled by the corresponding chemical potentials. However, strangeness conservation
is always treated  canonically. We consider three different models:
\begin{enumerate}
\item[(a)] the strangeness suppression, at fixed $T$ and $\mu_B$, is controlled by the volume
parameter which coincides with the system size,
\item[(b)] an additional strangeness suppression is introduced through  the $\gamma_S$-fugacity factor, \item[(c)] the canonical suppression is controlled by the cluster volume which { can be} smaller than the system size.
\end{enumerate}

From the comparisons shown in Fig.~\ref{fig2} it is clear the model (a) fails to describe the $p-p$
data. Although,   the canonical description of strangeness production is included in this model, the
strange particle ratios exhibit  large  deviations from data. The  discrepancies seen in
Fig.~\ref{fig2}  for model (a) could be even larger for multistrange particles. The reason of these
discrepancies  is due to the fact, that the volume of the system,  fixed from  the pion yields, is
already too large to imply   strangeness  suppression. From the $\chi^2$ fit to { SPS} data one
gets, in model (a) the system size radius  $R\simeq 1.3$. Consequently, as seen in Fig.~\ref{fig2},
at  this value of $R$  the canonical suppression of single-strange particles is indeed negligible.
Thus, for $S=\pm 1$ particles the  model (a) is almost equivalent to the GC treatment of strangeness
conservation.

From the above discussion it is clear, that at the SPS the   strangeness suppression due to
canonical effects alone  is not sufficient to describe the $4\pi$ data. In models (b) and (c) we
have included an additional suppression  of strange particle phase-space by introducing either the
$\gamma_S$ factor or the strangeness correlation volume. From the comparisons of the model  with
data shown in Fig.~\ref{fig2} one sees that both these models describe the { SPS} data quite well
with similar values for $T$ and $\mu_B$  (see Table~\ref{Table_fit}). The particle yields calculated
in these models are summarized in Table~\ref{Table_sps}. The only essential difference
between model (b) and (c) appears in  the yield of $\phi$ meson which in both cases is inconsistent
with data  as seen in Table~\ref{Table_sps} \footnote{The statistical model results  for hidden
strange particles are known to be inconsistent with data, particularly for elementary collisions.
For more detailed discussion see e.g. Ref.~\cite{small}.}, for a detailed discussion see Ref.~\cite{small}.
\input{table_fit.tex}
\input{table_sps.tex}
\begin{figure}
\includegraphics[width=0.99\linewidth]{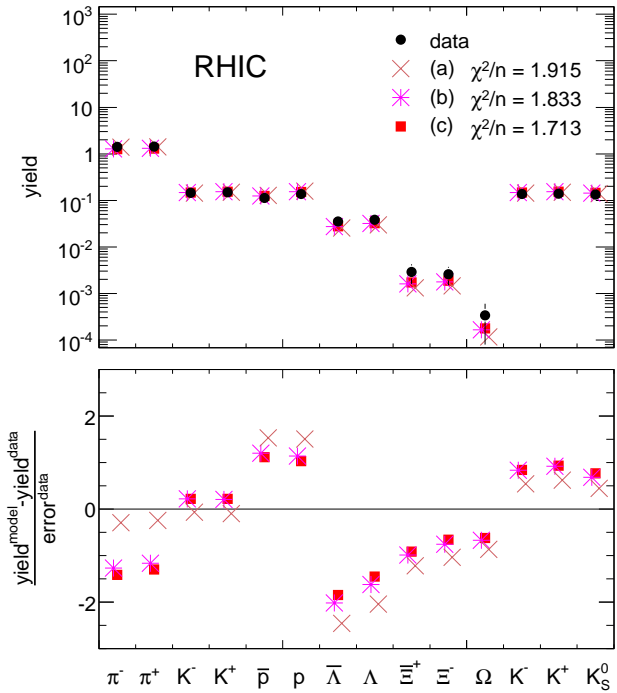}
\caption{\label{fig3} Midrapidity particle densities from p-p collisions at $\sqrt{s}$~=~200~GeV
from STAR~\cite{dataRHICstable,dataRHIChyp} together with the model fits as in Fig.~\ref{fig2}. The
lower panel shows the $\chi^2$-deviations of the model fits to data.}
\end{figure}

In order to quantify the change of thermal parameters with collision  energy  we also analyze  data
on particle production in $p-p$ collisions  at RHIC. Here, particle yields are only available around
midrapidity~\cite{dataRHICstable,dataRHIChyp}. However, contrary to SPS the yields of multistrange
particles are also available allowing for more complete verification of thermal  models used in the
analysis at SPS. In Fig.~\ref{fig3} we compare the predictions of models (a), (b) and (c) with RHIC
data. The particle yields used in this study and the yields calculated in the model are summarized
in  Table~\ref{Table_rhic}. The resulting thermal parameters for models (b) and (c) are indicated in
Table~\ref{Table_fit}. The STAR Collaboration published the statistical model analysis of  $p-p$ data
restricted to  yields of  pions, kaons and (anti)protons~\cite{dataRHICstable}, using the
undersaturation factor $\gamma_S$. Our analysis agrees  with Ref.~\cite{dataRHICstable} if one uses
the same data set. With the   hyperons included in the analysis there is   a slight increase in
$\gamma_S$ whereas the temperature and chemical potential stay almost the same.
\input{table_rhic.tex}

From Fig.~\ref{fig3} and Table~\ref{Table_rhic} one sees that all models describe RHIC data within
two standard deviations. At RHIC the pion yield at midrapidity is by more than a factor of
two lower than at SPS in $4\pi$, resulting in the  corresponding decrease of the volume parameter
and stronger strangeness suppression. In the context of  the considered RHIC data it is rather
difficult to definitely verify,  that there is a need for an extra strangeness suppression effects
going beyond the one already included in the standard canonical model. This is clear from
Table~\ref{Table_rhic} where in model (b) the $\gamma_S\sim 1$ and in model $(c)$ the cluster and
the system volume coincides within errors.

In the following, we concentrate on the statistical model predictions for particle yields at LHC
energy. In view of the results obtained at  SPS and at RHIC, we limit our attention only to model
(c) and its extrapolation to LHC energies.
\section{\label{secPredict}Particle ratios in $p-p$ collisions at LHC}
\begin{figure}
\includegraphics[width=0.99\linewidth]{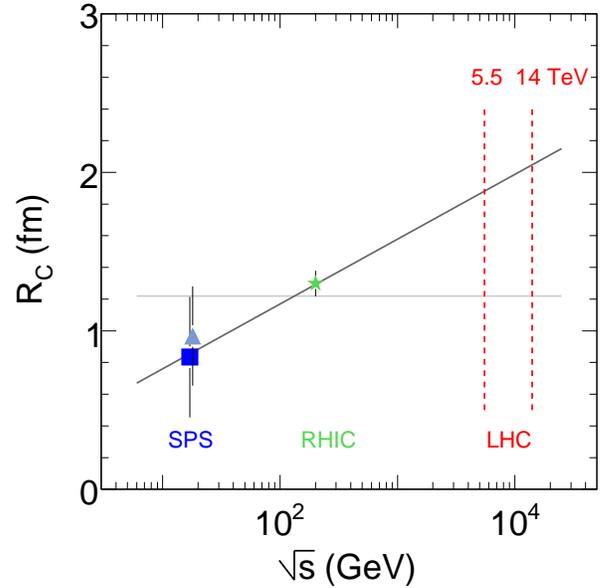}
\caption{\label{fig4} Cluster radius $R_C$ as a function of energy, extracted from the present analysis of SPS data (triangle) and RHIC data (star). Also shown is the previous analysis~\cite{small} of a smaller set of SPS data (square). The lines illustrate possible evolutions towards LHC energies as discussed in the text.}
\end{figure}
The extrapolation of particle ratios  to LHC energy requires  estimates of  the temperature, the
chemical potential and the cluster volume. From our analysis made at SPS and RHIC energies
(summarized in Tables  \ref{Table_sps} and \ref{Table_rhic}) it is clear that no variation in the
temperature is expected  between SPS, RHIC and LHC. A strong decrease of $\mu_B$ from SPS to RHIC
seen in our results, together with the  previous systematics on the beam energy dependence of
$\mu_B$ obtained from freezeout conditions in heavy-ion collisions \cite{wheaton_phd},  indicate
that all chemical potentials should be very small  at LHC. In the following,  we use $T\simeq 170$
MeV and $\mu_B\simeq 1$ MeV from Ref.~\cite{lhcPbPb} as appropriate thermal parameters at LHC.  This
temperature can be interpreted as the critical temperature  in the QCD phase transition and/or as
the Hagedorn limiting temperature of the hadronic medium.

The only parameter that remains largely uncertain in extrapolating  from SPS and RHIC to LHC is the size of
the cluster described by the radius  $R_C$ quantifying the strangeness suppression. Two
limiting cases for the extrapolation are used in the model  and shown in Fig.~\ref{fig4}:
\begin{description}
\item[i)] Saturation of the correlation radius at $R_C\simeq 1$ fm,
\item[ii)] An increase of $R_C$ from SPS and RHIC to LHC.
\end{description}
\begin{figure}
\includegraphics[width=0.99\linewidth]{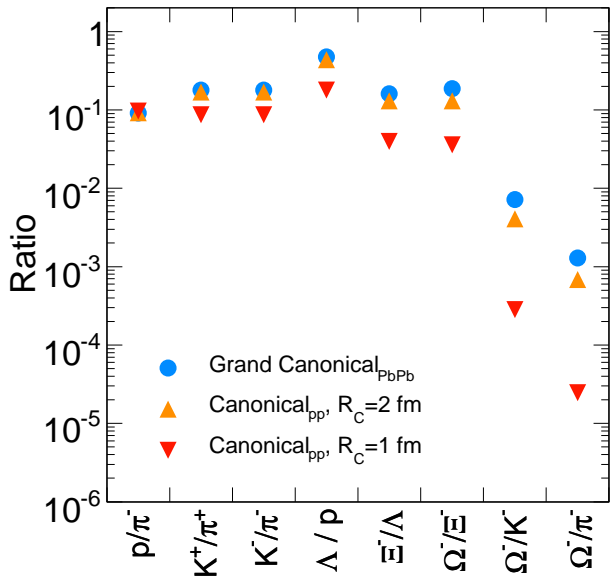}
\caption{\label{fig5} Plot Predictions for various particle ratios using different values for $R_C$.}
\end{figure}

In the first case, namely of an energy independent $R_C$ and at fixed temperature, the strangeness
suppression at LHC will be the same as at  RHIC. Consequently, different ratios of strange to
non-strange particle yields will be  modified only  through the variation in $\mu_B$ which can
be quantified by the $\exp(\pm\mu_B/T)$ Boltzmann factor.

In the second  case, of increasing $R_C$ with $\sqrt{s}$ and at fixed temperature, the
strangeness suppression at LHC will be weaker than at RHIC leading possibly to an equilibrated, canonical system without any additional suppression. The scenario  $ii)$
is naturally expected if $R_C$  coincides with the size of the system. In this case, the  $R_C$
should scale with the number of pions in the final state. This scenario could be verified
experimentally at LHC by comparing the strange/non-strange particle ratios in $p-p$ collisions for
events with different pion multiplicities. If valid, then for sufficiently high pion multiplicity at large $\sqrt{s}$ the strangeness production normalized to pion multiplicities could  converge to the results expected in heavy-ion collisions. However, in view of the known data in elementary and heavy-ion collisions this is very unlikely scenario. An increase of  $R_C$ with $\sqrt{s}$ shown in Fig.~\ref{fig4} is
expected to saturate at higher energies, or to be much weaker than linear. Due to lack of data the
actual dependence of $R_C=R_C(\sqrt s)$ is not known. The LHC data are essential to understand this
behavior of strangeness suppression and its energy dependence.
\begin{figure}
\includegraphics[width=0.99\linewidth]{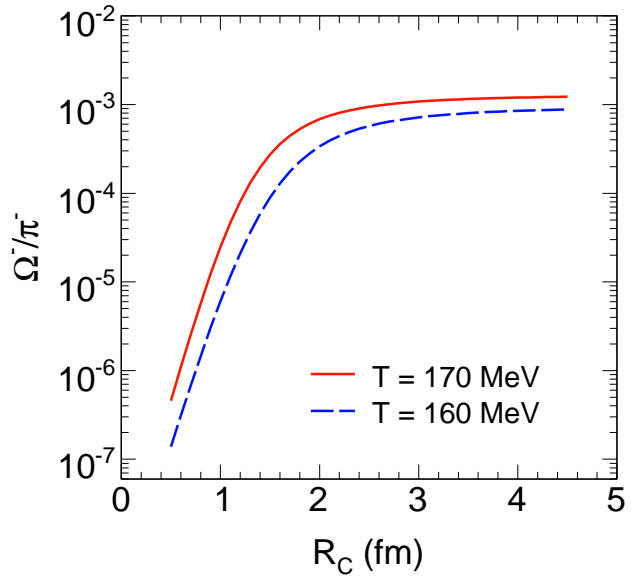}
\caption{\label{fig6} Particle ratio $\Omega/\pi$ as a function of $R_C$ for two assumed
temperatures, T = 160 MeV (dashed) and T = 170 MeV (full line). }
\end{figure}

In Fig.~\ref{fig5} and Table~\ref{Table_pred} we summarize predictions of the statistical model
for different particle ratios at LHC energy. We compare the results obtained under the GC
description of strangeness conservation with the canonical model (c) with the parameter $R_C$
describing the size where strangeness is conserved exactly in a system. Changing the value of $R_C$
from one to two fermi implies dramatic change in the ratios involving multistrange particles (see
Fig.~\ref{fig5}). It is interesting to note, that even for $R_C=2$ fm the $\Xi/\pi$ and $\Omega/\pi$
ratios differs from their GC values, whereas $K/\pi$ and $\Lambda/\pi$ are already well consistent
with GC results. The $\Omega/\pi$  is particularly   sensitive to changes  in $R_C$  while its
temperature dependence is rather moderate as seen in Figures~\ref{fig5} and \ref{fig6}.
Consequently, the $\Omega/\pi$ ratio is an excellent observable to probe strangeness suppression and
correlated strangeness production in $p-p$ collisions at LHC energy.
\input{table_pred}
\section{\label{secSummary}Summary}
We have analyzed, within the statistical model,   recent data on particle production  in
$p-p$ collisions at  SPS and RHIC energies. The models were formulated in the canonical ensemble with
respect to strangeness conservation and extended to implement extra strangeness suppression
either by  the  off-equilibrium $\gamma_S$  or by a  strangeness correlation volume. We have shown, that at
RHIC the midrapidity data can be successfully described by the canonical statistical model with and
without any  extra suppression effects.
On the other hand, the full phase-space SPS data require additional strangeness suppression  that is
beyond the conventional canonical model.  Extrapolating all relevant thermal parameters from
SPS and  RHIC up to LHC energy we have made  predictions  for particle yields in $p-p$ collisions
at $\sqrt{s}$ = 14~TeV. We have discussed the role and the influence of strangeness correlation
volume on particle production in p-p collisions. We have argued that the $\Omega/\pi$ ratio is an
excellent probe of strangeness correlations and/or strangeness suppression mechanism in $p-p$
collisions. We have indicated that comparing  strangeness production at the LHC in events with
different pion multiplicities can provide a deep insights into  our understanding of strangeness
suppression from $A-A$ to $p-p$ collisions.
\begin{acknowledgments}
One of us (K.R.) acknowledges support from  DFG-Mercator program. Two of us (J.C. and H.O.)
acknowledge support of the   DFG-NRF grant. The work of K.R. and J.C. was partly  supported by the
research project within South Africa -  Poland scientific collaboration.
\end{acknowledgments}
\end{document}

%% file: table_fit.tex
\begin{table}
% \begin{center}
\caption{\label{Table_fit} Statistical model parameters, extracted from the comparison of model (c) (upper table) and model (b) (lower table) with experimental data on p-p collisions. Full phase-space (4$\pi$ integrated) data at $\sqrt{s}$~=~17.3~GeV and midrapidity densities at $\sqrt{s}$~=~200~GeV were used.}
\begin{ruledtabular}
\begin{tabular}{lcccc}
$\sqrt{s} \rm{(GeV)}$ & $T \rm{(MeV)}$ &  $\mu_B \rm{(MeV)}$ &  $R \rm{(fm)}$ &  $R_C \rm{(fm)}$ \\
\hline
17.3&169$\pm$16&225$\pm$45&1.4$\pm$0.3&1.0$\pm$0.3\\
200&165$\pm$4&14$\pm$18&1.2$\pm$0.1&1.3$\pm$0.1\\
%
% \hline
\\
% \hline
% 
$\sqrt{s} \rm{(GeV)}$ & $T \rm{(MeV)}$ &  $\mu_B \rm{(MeV)}$ &  $R \rm{(fm)}$ &  $\gamma_S$ \\
\hline
17.3&164$\pm$9&211$\pm$27&1.5$\pm$0.2&0.7$\pm$0.1\\
200&165$\pm$4&14$\pm$18&1.2$\pm$0.1&1.1$\pm$0.1\\
\end{tabular}
\end{ruledtabular}
\end{table}

%% file: table_sps.tex
\begin{table}
% \begin{center}
\caption{\label{Table_sps}
Particle yields (4$\pi$ integrated) in minimum bias p-p collisions at $\sqrt{s}$~=~17.3~GeV from Ref.~\cite{dataSPSpi,dataSPSK,dataSPSlam,dataSPSK0s} and 
fit results from models (b) and (c).
The $\phi$ yield (below the line, Ref.~\cite{dataSPSphi}) was not included in the fit. The values given here are model predictions using the thermal parameters extracted from the analysis of the remaining data.}
\begin{ruledtabular}
\begin{tabular}{lccc}
Particle &Data&Fit (b)&Fit (c)\\
\hline
$\pi^{-}$&3.02 $\pm$ 0.15 & 3.01 & 3.01 \\
$\pi^{+}$&2.36 $\pm$ 0.11 & 2.32 & 2.32 \\
$K^{-}$&0.258 $\pm$ 0.055 & 0.268 & 0.270 \\
$K^{+}$&0.160 $\pm$ 0.050 & 0.162 & 0.166 \\
$\Lambda$&0.116 $\pm$ 0.011 & 0.119 & 0.120 \\
$\bar{\Lambda}$&0.0137 $\pm$ 0.0007 & 0.0135 & 0.0134 \\
$K^{0}_{S}$&0.18 $\pm$ 0.04 & 0.21 & 0.20 \\
$\bar{p}$&0.0400 $\pm$ 0.0068 & 0.0464 & 0.0469 \\
\hline
$\phi$&0.0120 $\pm$ 0.0015 & 0.0271 & 0.0598 \\
\end{tabular}
\end{ruledtabular}
\end{table}

%% file: table_rhic.tex
\begin{table}
% \begin{center}
\caption{\label{Table_rhic}
Particle midrapidity densities in minimum bias p-p collisions at $\sqrt{s}$~=~200~GeV from Ref.~\cite{dataRHICstable,dataRHIChyp} and fit results from models (b) and (c).
Charged kaons were analysed with two different techniques and both measurements are included in the fits.
The $\phi$ yield (below the line, Ref.~\cite{dataRHICphi}) was not included in the fit. The values given here are model predictions using the thermal parameters extracted from the analysis of the remaining data.}
\begin{ruledtabular}
\begin{tabular}{lccc}
Particle &Data&Fit (b)&Fit (c)\\
\hline
$\pi^{-}$&1.42 $\pm$ 0.11 & 1.28 & 1.26 \\
$\pi^{+}$&1.44 $\pm$ 0.11 & 1.31 & 1.30 \\
$K^{-}$&0.145 $\pm$ 0.013 & 0.148 & 0.148 \\
$K^{+}$&0.150 $\pm$ 0.014 & 0.153 & 0.153 \\
$\bar{p}$&0.113 $\pm$ 0.010 & 0.125 & 0.124 \\
$p$&0.138 $\pm$ 0.013 & 0.153 & 0.151 \\
$\bar{\Lambda}$&0.0351 $\pm$ 0.0039 & 0.0272 & 0.0279 \\
$\Lambda$&0.0385 $\pm$ 0.0042 & 0.0317 & 0.0324 \\
$\bar{\Xi}^{+}$&0.0029 $\pm$ 0.0013 & 0.0016 & 0.0017 \\
$\Xi^{-}$&0.0026 $\pm$ 0.0011 & 0.0018 & 0.0019 \\
$\Omega^{-} + \bar{\Omega}^{+}$&0.00034 $\pm$ 0.00026 & 0.00017 & 0.00018 \\
$K^{-}$ (kink)&0.137 $\pm$ 0.013 & 0.148 & 0.148 \\
$K^{+}$ (kink)&0.140 $\pm$ 0.014 & 0.153 & 0.153 \\
$K^{0}_{S}$&0.134 $\pm$ 0.014 & 0.144 & 0.145 \\
\hline
$\phi$&0.0180 $\pm$ 0.0037 & 0.0391 & 0.0317 \\
\end{tabular}
\end{ruledtabular}
\end{table}

%% file: table_pred.tex
\begin{table}
% \begin{center}
\caption{\label{Table_pred} Particle ratios for thermal conditions expected 
at LHC. The extreme values of the cluster size $R_C$ span the band of expected 
numerical values. 
The grand-canonical values in the last column are included 
for comparison, for a detailed discussion see Ref.~\cite{lhcPbPb}.}
\begin{ruledtabular}
\begin{tabular}{lccc}
Ratio & $R_C$ = 1 fm & $R_C$ = 2 fm & grand canon. \\
\hline
$p / \pi^-$ 		& 0.0970 	& 0.0920 & 0.0914 \\
$K^+ / \pi^+$ 		& 0.0871 	& 0.169 & 0.180 \\
$K^- / \pi^-$ 		& 0.0870 	& 0.169 & 0.179 \\
$\Lambda / p$ 		& 0.179 	& 0.436 & 0.473 \\
$\Xi^- / \Lambda$ 	& 0.0397 	& 0.130 & 0.160 \\
$\Omega^- / \Xi^-$ 	& 0.0358 	& 0.131 & 0.186 \\
$\Omega^- / K^-$ 	& 2.83 $\cdot ~ \rm 10^{-4}$ & 4.06 $\cdot ~ \rm 10^{-3}$ & 7.19 $\cdot ~ \rm 10^{-3}$ \\
$\Omega^- / \pi^-$ 	& 2.46 $\cdot ~ \rm 10^{-5}$ & 6.85 $\cdot ~ \rm 10^{-4}$ & 1.29 $\cdot ~ \rm 10^{-3}$ \\
\end{tabular}
\end{ruledtabular}
\end{table}